\documentclass[runningheads]{llncs}
\usepackage[T1]{fontenc}
\usepackage{graphicx}
\usepackage{amsmath}
\usepackage{orcidlink}
\usepackage{caption}
\usepackage{subcaption}
\usepackage{booktabs}

\begin{document}
\title{A Gradient-Optimized TSK Fuzzy Framework for Explainable Phishing Detection}
\titlerunning{Explainable Phishing Detection}

\author{Lohith Srikanth Pentapalli\,\orcidlink{0009-0002-9801-3418}\inst{1} \and Jon Salisbury\inst{2} \and Josette Riep\orcidlink{0000-0002-7325-4381}\inst{1} \and Kelly Cohen\,\orcidlink{0000-0002-8655-1465}\inst{1}}
\authorrunning{Pentapalli L.S., et al.}

\institute{University of Cincinnati \and Nexigen \\
\email{pentapls@mail.uc.edu, jsalisbury@nexigen.com, riepjr@ucmail.uc.edu, cohenky@ucmail.uc.edu}}

\maketitle

\begin{abstract}
    Phishing attacks represent an increasingly sophisticated and pervasive threat to individuals and organizations, causing significant financial losses, identity theft, and severe damage to institutional reputations. Existing phishing detection methods often struggle to simultaneously achieve high accuracy and explainability, either failing to detect novel attacks or operating as opaque black-box models. To address this critical gap, we propose a novel phishing URL detection system based on a first-order Takagi-Sugeno-Kang (TSK) fuzzy inference model optimized through gradient-based techniques. Our approach intelligently combines the interpretability and human-like reasoning capabilities of fuzzy logic with the precision and adaptability provided by gradient optimization methods, specifically leveraging the Adam optimizer for efficient parameter tuning. Experiments conducted using a comprehensive dataset of over 235,000 URLs demonstrate rapid convergence, and exceptional predictive performance (accuracy averaging 99.95\% across 5 cross-validation folds, with a perfect AUC i.e. 1.00). Furthermore, optimized fuzzy rules and membership functions improve interpretability, clearly indicating how the model makes decisions -- an essential feature for cybersecurity applications. This high-performance, transparent, and interpretable phishing detection framework significantly advances current cybersecurity defenses, providing practitioners with accurate and explainable decision-making tools.
\end{abstract}

\section{Introduction}

Phishing is a widespread and evolving form of cybercrime that deceives individuals into disclosing confidential information such as passwords and financial data. The most prominent mediums for phishing are emails, social media, and SMS (smishing), where a malicious message that evokes urgency is sent to the victim. The message contains a malicious link that redirects the victim to a webpage similar to an authentic website from a reputable service \cite{alkhalilPhishingAttacksRecent2021}. As the digital landscape continues to grow, phishing represents a persistent and evolving threat to cybersecurity. Attackers often use social engineering techniques to exploit human psychology, creating a sense of urgency or fear to compel victims to act quickly and divulge sensitive information without verifying the authenticity of the request \cite{prajapatiSocialEngineering2024}.

Phishing techniques have significantly evolved in terms of prevalence and sophistication over time. The impact of phishing is profound and multifaceted. On an individual level, victims may suffer financial losses, identity theft, and emotional distress. The consequences can be even more severe for organizations as it can lead to data breaches, which can cause severe reputational damage and legal ramifications. Phishing is responsible for 36\% of data breaches in the US \cite{2024PhishingStatistics}. Phishing toolkits have also made it relatively cheap for malicious actors to launch phishing campaigns \cite{purwantoPhishSimAidingPhishing2022}, possibly leading to the rise in phishing attacks.

Due to the proliferation of sophisticated phishing campaigns, there is a growing need for detection systems that are accurate and explainable. Explainability ensures that the decisions made by a detection system can be understood, trusted, and validated by human analysts or stakeholders. This is particularly vital in cybersecurity, where opaque decisions from black-box models can hinder incident response, compliance, and the continuous refinement of defense strategies. Security teams should understand why the detection system classified a malicious URL as it did. This transparency fosters trust in the detection system, which is crucial for its widespread use - hence the motivation behind using fuzzy logic-based systems in phishing detection.

Fuzzy logic is a promising approach for handling the inherent uncertainty and imprecision that characterizes phishing detection. Binary logic systems operate on crisp classifications with rigid boundaries, whereas fuzzy logic reflects the nuances of human reasoning by allowing degrees of truth rather than absolute states \cite{zadehSoftComputingFuzzy1994}. This approach makes it extremely effective in phishing scenarios where different indicators – such as the lexical and HTML features of a URL and webpage, respectively – do not conform to a strict boundary but could have degrees of membership in different classes. The Takagi-Sugeno-Kang (TSK) fuzzy inference system, a powerful variant of fuzzy logic models, excels at modeling complex, nonlinear systems using interpretable if-then rules coupled with mathematical functions \cite{takagiFuzzyIdentificationSystems1985}. Its structure facilitates extracting meaningful patterns from ambiguous input data, which is invaluable in the context of phishing, where attackers constantly innovate to evade detection. However, while TSK systems are highly interpretable, they rely on some sort of expert knowledge to model the membership functions and the rules necessary to perform effective classification. Without expertise concerning highly evolved and sophisticated phishing attacks, we must incorporate some learning methodology to form a robust rule base. To address this, we integrate gradient-based optimization into the TSK framework, allowing it to fine-tune its parameters on data dynamically. This hybrid approach harnesses the best of both worlds – retaining the transparency and human-like reasoning of fuzzy logic while enhancing learning efficiency and adaptability through gradient descent. In this work, we propose a novel and intelligent phishing detection system that leverages this synergy to deliver high accuracy without sacrificing explainability, offering a highly effective, transparent, and forward-looking solution in the ongoing fight against increasingly sophisticated phishing attacks.

\section{Background}

Over the years, a variety of techniques have emerged for phishing detection. One of the earliest form of approaches involve blacklisting which even though is simple, often falls short against novel or rapidly changing phishing sites due to which the URL repositories that they depend on need to be updated regularly to be reliable \cite{prakashPhishNetPredictiveBlacklisting2010}\cite{abuadbbaWebPhishingDetection2022}. Therefore, it is not considered robust enough to be deployed in real-time systems or is only suitable as a first layer of defense. Heuristic-based techniques improved on this by analyzing webpage structures and URL patterns, offering a rule-driven perspective that could identify anomalies indicative of phishing \cite{khonjiPhishingDetectionLiterature2013}. However, these static rules suffer from scalability and accuracy issues \cite{abuadbbaWebPhishingDetection2022}. 

Recently, there has been a profound advent of machine learning for phishing detection \cite{safiSystematicLiteratureReview2023}. ML-based detection typically involves extracting a set of features (e.g., URL tokens, page element frequencies, domain reputation metrics) and training a classifier to distinguish phishing sites from legitimate ones \cite{abuadbbaWebPhishingDetection2022}. Common algorithms include Support Vector Machines, Random Forests, and logistic regression, which can generalize beyond specific known attacks by capturing statistical regularities in phishing websites. Unlike blacklists, ML models can potentially detect zero-day phishing pages by recognizing tell-tale features not present in benign sites \cite{abuadbbaWebPhishingDetection2022}. In fact, ML-based approaches have been widely adopted (even built into modern web browsers) due to their ability to scale and adapt to new phishing strategies \cite{abuadbbaWebPhishingDetection2022}. 

Deep learning approaches represent the latest generation of phishing detection techniques. Deep neural networks such as Convolutional Neural Networks (CNNs), Recurrent Neural Networks (RNNs), and transformer-based models have been applied to phishing detection in various forms. These models can automatically learn features from raw input (URL strings, email text, or screenshots of websites) and have shown excellent accuracy, often exceeding that of shallow classifiers \cite{songailaiteBERTBasedModelsPhishing}. For example, recurrent models such as LSTMs and BiLSTMs are among the most widely used deep learning approaches for phishing detection, capable of capturing sequence patterns in URLs or email content \cite{songailaiteBERTBasedModelsPhishing}. Studies have explored transformer-based architectures (BERT-based models) for phishing email classification and found to be achieving over 98\% accuracy, far outperforming a traditional classifier on the same task \cite{alhuzaliInDepthAnalysisPhishing2025}. However, this increase in accuracy comes at the cost of greater computational requirements and a lack of interpretability. Machine learning and deep learning models act as complex black boxes, making it unclear why a given site is flagged \cite{calzarossaAssessmentFrameworkExplainable2025}. This opacity poses challenges for security analysts and end-users, motivating research into more explainable solutions. 

In critical security applications, a detection system’s output often needs to be understood and verified – for instance, an analyst investigating an alert benefits from knowing which features were indicative of phishing. Moreover, regulators and organizational policies are increasingly calling for interpretable and transparent AI in critical applications such as healthcare and finance \cite{calzarossaAssessmentFrameworkExplainable2025}. Explainability is seen as essential for human oversight, allowing stakeholders to comprehend the reasons behind an alert and to ensure the system’s decisions are justified \cite{calzarossaAssessmentFrameworkExplainable2025}. In the phishing domain in particular, explainability is a compelling issue because of its potential benefits for detecting fraud and designing better defenses. An explainable phishing detector can provide insights into attackers’ tactics (e.g., which visual elements or URL tricks fooled the model), helping refine security measures and user education \cite{calzarossaAssessmentFrameworkExplainable2025}.

Cybersecurity has employed fuzzy logic to handle uncertainty and linguistic reasoning in detection systems. Unlike binary logic, a fuzzy logic system works with degrees of truth, allowing an element (e.g., a website or network event) to belong partially to multiple categories. This provides a natural way to draw conclusions from ambiguous or noisy inputs \cite{khraisatSurveyIntrusionDetection2019}. In intrusion detection, for example, the boundary between normal and malicious behaviour is often not sharply defined - a slight deviation from normal may or may not indicate an attack. Traditional binary rules or hard thresholds can therefore cause high false alarm rates (flagging benign anomalies) or missed detections \cite{khraisatSurveyIntrusionDetection2019}. Fuzzy logic addresses this by letting conditions be satisfied to a degree. An intrusion detection system can use fuzzy sets for features like CPU usage or packet rates (e.g., “high”, “medium”, “low”), and a fuzzy inference engine will combine them to output an alert level. Because an event can be, say, 70\% “suspicious” and 30\% “normal”, the system is more tolerant of minor deviations, which keeps false positive rates low while still catching serious anomalies \cite{khraisatSurveyIntrusionDetection2019}. 

The strength of fuzzy logic lies in encoding expert knowledge as if-then rules and handling imprecise data gracefully. A fuzzy phishing detector might use rules such as “IF URL has many hex characters (high) AND page has login form (yes) THEN phishing likelihood = high,” with each condition measured in fuzzy terms (e.g., “many” could be a fuzzy set on the number of hexadecimal characters in the URL). Such a system can integrate multiple soft signals to reach a decision. Studies indeed show that fuzzy rule-based classifiers can achieve high detection rates while minimizing false alerts by smoothing the boundary between benign and malicious content \cite{almseidinCyberPhishingWebsiteDetection2022}. For example, Almseidin et al. (2022) used fuzzy rule interpolation for phishing website detection and reported a 97.6\% detection rate with significantly reduced false positives, thanks to the fuzzy system’s ability to interpolate between known rules and handle uncertainty in the input \cite{almseidinCyberPhishingWebsiteDetection2022}. This illustrates how fuzzy logic’s “gray area” reasoning is well-suited to security problems where strict binary logic falters.

Takagi-Sugeno-Kang (TSK) fuzzy systems are a specific type of fuzzy inference model distinguished by their use of crisp outputs in the rule consequents. In a TSK system, each fuzzy rule has an output defined by a mathematical function (often a linear combination of the inputs, or a constant) rather than a linguistic term. For example, a TSK rule might be: IF URL-length is High AND Domain-Age is Low THEN Phish-Risk = 0.8, where 0.8 is a crisp value indicating risk. During inference, all applicable rules fire to a certain degree and the TSK model computes a weighted average of the rule outputs, yielding a final numerical score or decision. This design means that “in such systems consequents are functions of inputs” \cite{schererTakagiSugenoFuzzySystems2012}, and the output is obtained by a weighted interpolation of those functions \cite{takagiFuzzyIdentificationSystems1985}, \cite{schererTakagiSugenoFuzzySystems2012}. TSK fuzzy systems have proven to be powerful and convenient, especially for control and approximation tasks. Takagi and Sugeno’s seminal work showed that fuzzy rule systems of this type can be applied to industrial processes such as the water-cleaning process and a converter in steel-making process \cite{takagiFuzzyIdentificationSystems1985}. Additionally, they have also been proven to be universal approximators for various multivariate continuous functions \cite{yingGeneralTakagiSugenoFuzzy1998}. 

To fully realize the potential of fuzzy systems in complex domains, researchers have incorporated various optimization techniques to automatically learn or tune fuzzy rules and membership functions. Genetic Algorithms (GAs) are one prevalent approach for optimizing fuzzy systems. GAs can optimize both the structure and parameters of a fuzzy system \cite{fernandezEvolutionaryFuzzySystems2019} – for example, selecting which rules to include and fine-tuning the membership function shapes to maximize detection accuracy on training data. As a global search method, a GA can find near-optimal solutions even when the problem landscape is multimodal or discontinuous. The downside is that evolutionary searches can be computationally intensive; evaluating many candidate fuzzy systems against data is time-consuming, and convergence may require a large number of generations. 

In contrast to evolutionary methods, gradient-based optimization provides a more direct way to train fuzzy systems when they are formulated to be differentiable. The Adam optimizer has become a cornerstone in optimization due to key factors such as automatically scaling the gradients to handle parameters with varying magnitudes and its effective handling of sparse or noisy gradients \cite{kingmaAdamMethodStochastic2017}. Adam also requires less intensive tuning and remains stable across a wider range of learning rates when compared to the Stochastic Gradient Descent (SGD) \cite{sivaprasadOptimizerBenchmarkingNeeds2020}. It also offers practical advantages such as its ability to work effectively with mini-batch training and linear scaling with the number of parameters while maintaining a low memory footprint \cite{kingmaAdamMethodStochastic2017}. Due to its effectiveness for convex and non-convex optimization landscapes \cite{kingmaAdamMethodStochastic2017}, and due to its balance of speed, stability, and ease of use \cite{sivaprasadOptimizerBenchmarkingNeeds2020}, this makes an ideal choice for an optimizer.

In summary, current phishing detectors either cannot detect novel attacks (if using blacklists/heuristics) or cannot explain their decisions (if using complex ML/DL methods). The proposed approach addresses this gap by leveraging a TSK fuzzy system – which inherently provides interpretability through its rule base – and optimizing it with Adam to maximize accuracy. This combination promises a new level of high-performance phishing detection with explainability, directly tackling the trade-off that has defined the state of the art so far. The remainder of this paper will detail the development of such a model and demonstrate how it meets the twin objectives of effectiveness and interpretability in phishing detection.

\section{Methodology} 

This section details our approach to developing a first-order Takagi-Sugeno-Kang (TSK) fuzzy inference system for phishing URL detection, optimized using gradient-based techniques. 

\subsection{Data Collection and Preprocessing}
Our experiments utilized a comprehensive phishing URL dataset collected from UCI which contains 235,795 records with 54 features which indicate the lexical features of the URL and the HTML features of the webpages \cite{prasadPhiUSIILDiverseSecurity2024}. The dataset includes 100,945 phishing URLs and 134,850 legitimate URLs. 

\begin{figure}[!htbp]
    \centering
    \includegraphics[width=0.8\linewidth]{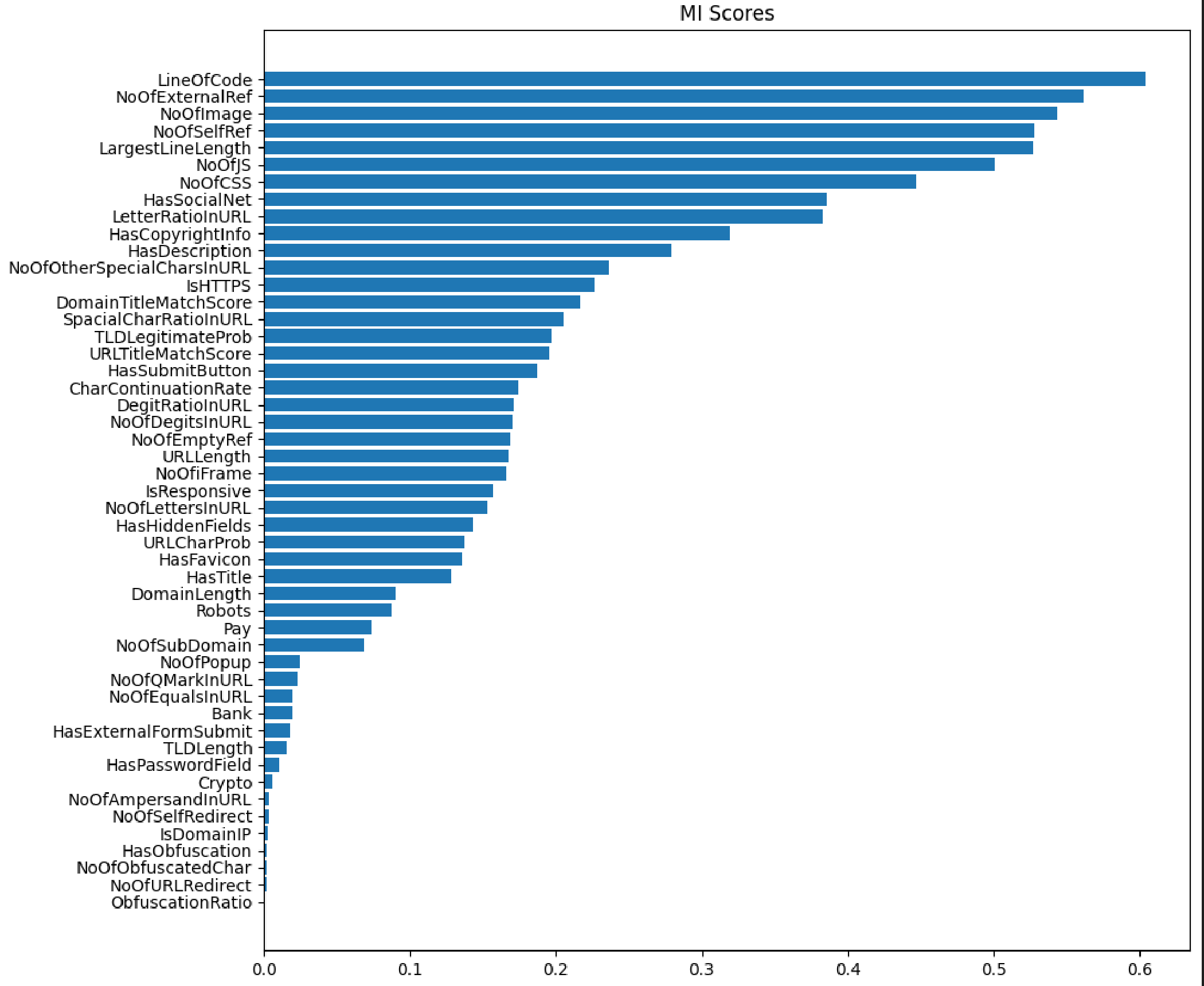}
    \caption{Mutual Information Scores}
    \label{mutual_info}
\end{figure}

Feature selection is crucial as including redundant features could degrade the performance of the system by introducing noise and increasing training time. Having a lot of features also inhibits the interpretability of the system since there would be an increase in the antecedent and consequent variables of the fuzzy system. We employed mutual information analysis to identify the most informative features for phishing detection. Mutual information helps in identifying features with high predictive relevance, greatly enhancing efficiency and interpretability by focusing on influential variables only. The mutual information scores for the features in our dataset are shown in Fig.~\ref{mutual_info}.

The feature "URLSimilarityIndex" was excluded as it serves as a strong proxy signal for legitimacy, potentially leaking target-label correlations into the feature space, which may artifically inflate performance metrics, while obscuring the model's ability to learn generalizable patters from lexical and HTML features of the URL and webpage respectively. The data was split into training (80\%) and testing (20\%) sets using stratified sampling to maintain class distribution. All features were standardized using z-score normalization to have zero mean and unit variance, ensuring that features with different scales contribute equally to the model.

\subsection{System Architecture}
The proposed system utilizes a first-order Takagi-Sugeno-Kang (TSK) fuzzy inference model with differentiable parameters, enabling end-to-end gradient optimization. Each feature is represented as a collection of Gaussian membership functions defined by the center (mean) and width (standard deviation) parameters. Gaussian membership functions provide smooth differentiability, which is essential for gradient-based methods, while also maintaining computational simplicity. Each input feature is then fuzzified using the Gaussian MFs. For each feature $x_i$, the membership degree $\mu_{A_{ij}}(x_i)$ to the $j$-th fuzzy set is calculated as:  

\begin{equation}
    \mu_{A_{ij}}(x_i) = \exp\left({-\frac{(x_i - c_{ij})^2}{2\sigma^2}}\right)
\end{equation}

The rule firing strength of each rule is calculated using the product t-norm operator. For rule $R_k$, the firing strength $\alpha_k$ is: 
\begin{equation}
\alpha_k = \prod_{i=1}^n \mu_{A_{i k_i}(x_i)}
\end{equation}
where  $k_i$ indicates the MF of feature $i$ used in rule $k$. 

Each rule has a linear consequent function in the following form:
\begin{equation}
    f_k(x) = b_k + \sum_{i=1}^n w_{ki}x_i
\end{equation}
where $b_{k}$ is the bias term and $w_{ki}$ are the linear coefficients for rule $k$

The final output is computed as the weighted average of rule consequents:
\begin{equation}
    y = \frac{\sum_{k=1}^m \alpha_k f_k(x)}{\sum_{k=1}^m \alpha_k}
\end{equation}
where $m$ is the number of rules.

A sigmoid function transforms continuous outputs into probabilities, suitable for binary classification tasks. The sigmoid ensures that predictions are probabilistically interpretable, directly mapping output values into a meaningful scale. Therefore, for the binary phishing detection task, the output is passed through a sigmoid function:
\begin{equation}
    p = \sigma(y) = \frac{1}{1 + e^{-y}}
\end{equation}

By formulating the architecture in this way, we can ensure interpretable rule-based reasoning while making the model flexible for gradient-based optimization. The system parameters include the Gaussian MF parameters (centers and widths) and the consequent parameters (linear coefficients and biases), all of which are optimized during training.

Each rule in the system takes the form: \\ \\
IF $x_1$ is $A_{1j}$ AND $x_2$ is $A_{2j}$ AND ... AND $x_n$ is $A_{nj}$ THEN $y = b_j + w_{j1}x_1 + w_{j2}x_2 + ... + w_{jn}x_n$ \\ \\
where $A_{ij}$ represents the fuzzy set for feature $i$ in rule $j$, and $b_j, w_{ji}$ are the consequent parameters for rule $j$.

The centers and width of the Gaussian MFs are decided by performing k-means clustering on the feature values, where the number of clusters per feature is decided by the number of MFs used to represent the feature. The standard deviation of membership functions is computed by determining the spread of points in their respective clusters. By doing so, we ensure a warm start instead of randomly initializing the parameters of the MFs, which leads to a significant reduction in the training time and faster convergence to the optimal solution. For binary features, we create tight Gaussian MFs that are centered at each unique value. The consequent parameters are initialized randomly to values between -1 and 1.  

\subsection{Parameter Optimization}

We implement a gradient-based optimization approach to tune all parameters of the TSK fuzzy system which includes the centers ($c_{ij}$) and widths ($\sigma_{ij})$ of the Gaussian membership functions and the linear coefficients ($w_{ji}$) and bias terms ($b_j$) of each rule.

The optimization is formulated as:
\begin{equation}
\min_{\theta} \mathcal{L}(\theta) 
= \frac{1}{N} \sum_{k=1}^N \mathrm{BCE}\bigl(y_k, \hat{y}_k\bigr)
+ \lambda \,\|\theta\|_2^2
\end{equation}
where:
\begin{itemize}
    \item $\theta$ represents all parameters of the model
    \item $BCE$ is the binary cross-entropy loss
    \item $y_k$ is the true label for the $k$-th example
    \item $\hat{y}_k$ is the predicted probability from the model
    \item $\lambda$ is the regularization coefficient
    \item $||\theta||_2^2$ is the L2 regularization term
\end{itemize}

The Adam optimizer is used for parameter updates due to its effectiveness with noisy gradients and ability to adapt learning rates [19]. For each parameter $\theta_i$, the update rule is:
\begin{equation}
\theta_i^{(t+1)}
= \theta_i^{(t)}
- \frac{\eta}{\sqrt{\hat{v}_i^{(t)}} + \epsilon}
\, \hat{m}_i^{(t)}.
\end{equation}
where $\hat{m}_i^{(t)}$ and $\hat{v}_i^{(t)}$ are the bias-corrected first and second moment estimates of the gradient, and $\eta$ is the learning rate. To ensure model robustness, we incorporate early stopping to prevent overfitting.

\subsection{Evaluation Metrics}

To comprehensively evaluate the performance of our optimized first-order TSK fuzzy system for phishing detection, the following metrics were employed:
\begin{enumerate}
    \item Accuracy: The proportion of correctly classified instances.
    \item Precision: The proportion of true positive predictions among all positive predictions, measuring the model's ability to avoid false positives.
    \item Recall: The proportion of true positive predictions among all actual positives, measuring the model's ability to detect phishing URLs.
    \item F1 Score: The harmonic mean of precision and recall, which balances the trade-off between false positives and false negatives
    \item Area Under the ROC Curve (AUC): Measuring the model's ability to discriminate between classes across various threshold settings.
\end{enumerate}

\section{Results and Discussion}

This section presents the experimental outcomes of our proposed first-order TSK fuzzy inference system, where we focus on the training dynamics and convergence, membership function evaluation, rule interpretability and inference behaviour, feature and rule sensitivity analysis, model performance evaluation using k-fold validation and a benchmark comparison with traditional classifiers.
\begin{figure}
    \centering
    \includegraphics[width=\textwidth]{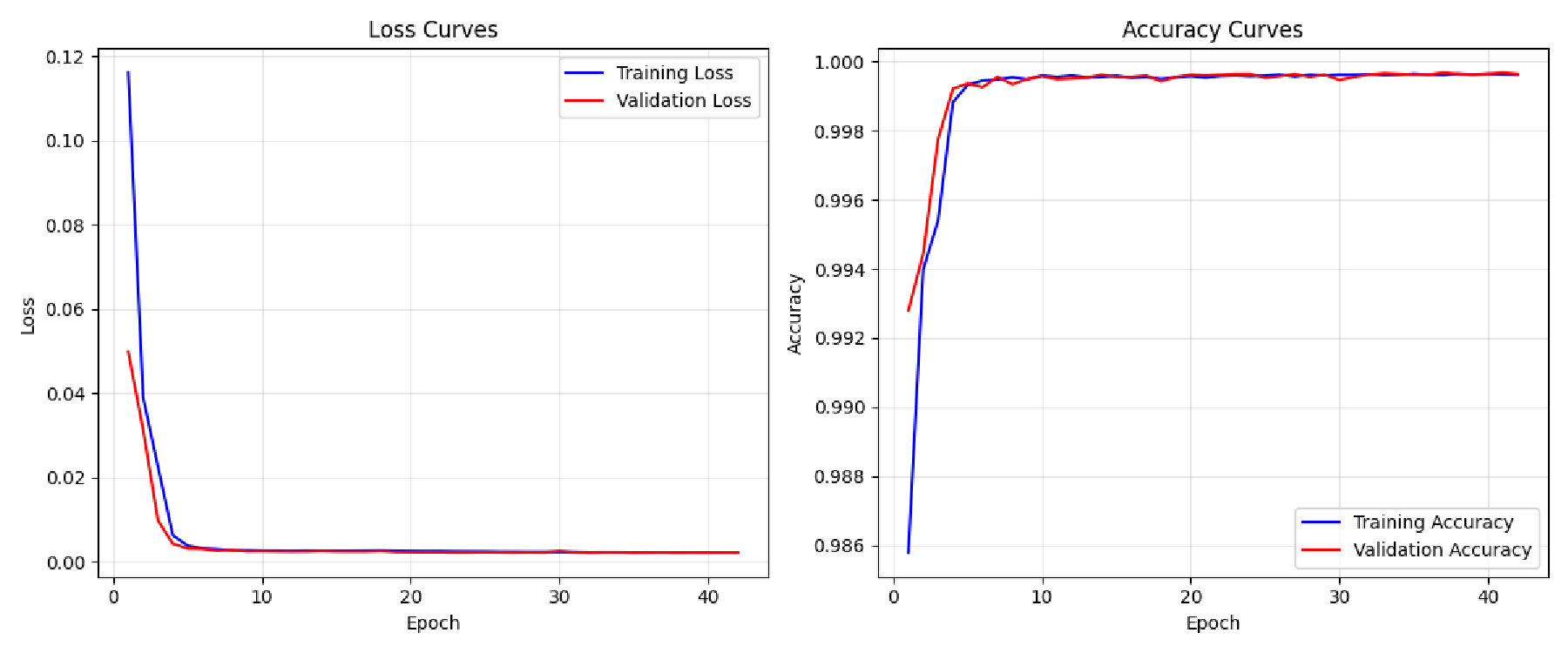}
    \caption{Training and Validation Loss Curves (left), Training and Validation Accuracy Curves (right)}
    \label{fig:loss-accuracy}
\end{figure}
For initial analysis, we selected the top 30 features according to their mutual information score, set the number of rules to 10, and set the number of membership functions of the continuous features to 3. The learning rate is set at 0.005. 

Figure \ref{fig:loss-accuracy} illustrates the loss and accuracy curves of our model over the course of training. The left plot shows the evolution of both training and validation losses, while the right plot depicts the progression of training and validation accuracies. One noteworthy observation from these curves is the model's ability to converge rapidly within the first 5 epochs. Within the first 5 epochs, both training and validation losses sharply decrease and then stabilize quite quickly thereafter. This demonstrates the Adam optimizer's effectiveness in converging at near-optimal solutions in the parameter space efficiently and rapidly. Furthermore, we observe a remarkably close alignment between the training and validation curves throughout the training process. The tightly matched trajectories of the loss curves indicate that the model is effectively generalizing beyond the training data, avoiding overfitting or underfitting. The accuracy curves further prove the model's strong performance, quickly surpassing 99.9\% accuracy on both the training and validation sets within the first few epochs. This high accuracy persists throughout the entire training duration, underscoring the robustness and effectiveness of our TSK fuzzy inference system in accurately classifying phishing URLs.

\begin{figure}
\centering
\begin{subfigure}{.5\textwidth}
  \centering
  \includegraphics[width=\linewidth]{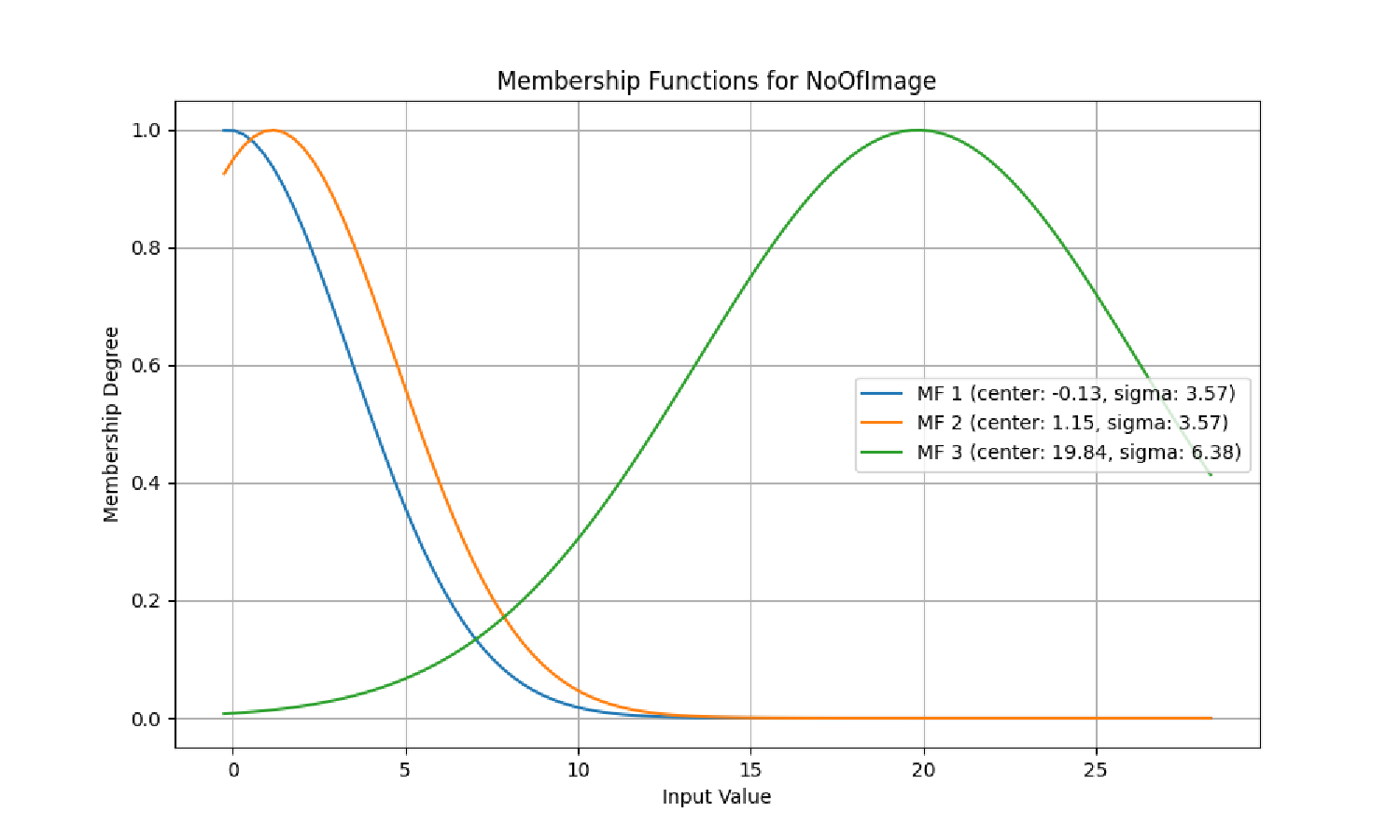}
  \caption{Before optimization}
  \label{fig:sub1}
\end{subfigure}%
\begin{subfigure}{.5\textwidth}
  \centering
  \includegraphics[width=\linewidth]{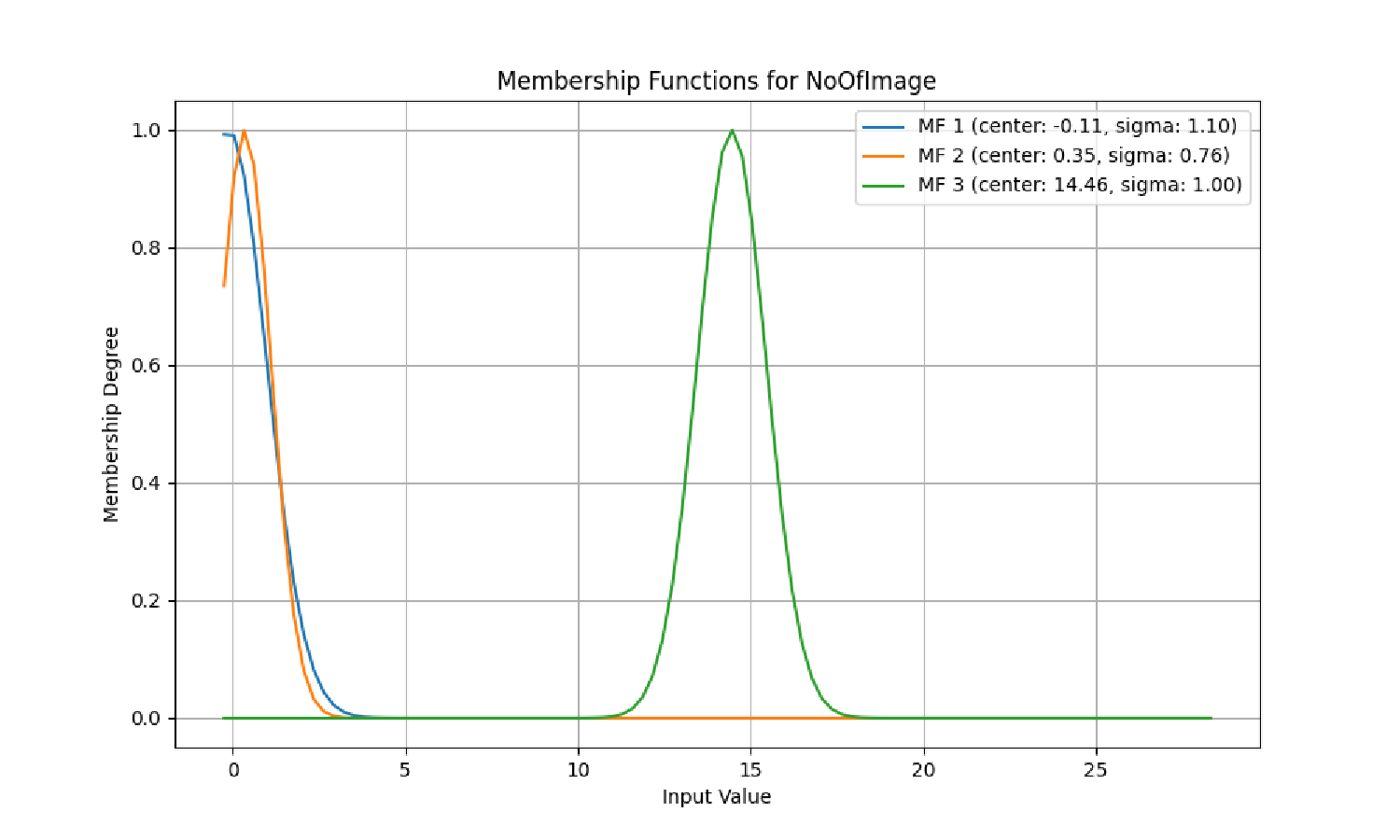}
  \caption{After optimization}
  \label{fig:sub2}
\end{subfigure}
\caption{MFs of feature "NoOfImages"}
\label{fig:mfs_before_after}
\end{figure}

Figure \ref{fig:mfs_before_after} illustrates how the Gaussian MFs for the feature "NoOfImages" evolved from their initial k-means based positions to their optimized final state. The MFs became significantly narrower, sharpening their definitions and reducing overlaps, with new centers located around -0.11, 0.35, and 14.46 from -0.13, 1.15, and 19.84, respectively. This underscores the model's ability to fine-tune itself, better reflecting the underlying data distribution and improving its discriminative performance.

\begin{figure}
    \centering
    \includegraphics[width=\textwidth]{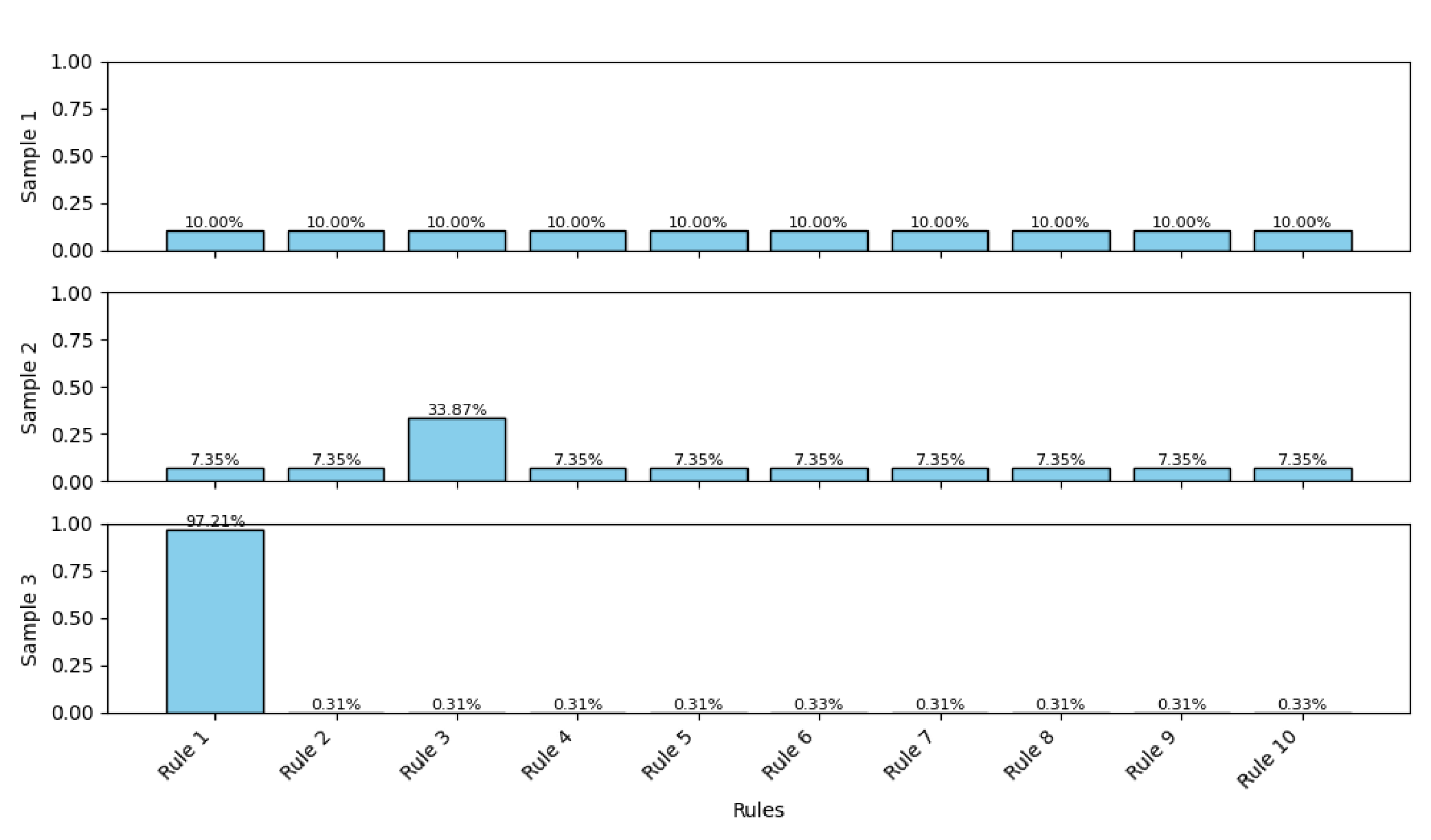}
    \caption{Rule activations for 3 samples in Test data}
    \label{fig:rule_activations}
\end{figure}

Figure \ref{fig:rule_activations} illustrates the activation patterns of fuzzy rules for 3 distinct samples from the test dataset, offering insights into how the model applies its rule base differently depending on the input characteristics. Sample 1 evenly engages all ten rules, whereas sample 2 predominantly activates rule 3 (33.87\%) while still partially engaging the other rules. This suggests a stronger but not exclusive alignment with the conditions described by Rule 3. Most strikingly, Sample 3 demonstrates remarkable rule specialization, with Rule 1 overwhelmingly activated at 97.21\%, effectively dominating the decision-making process. This strong specialization of Rule 1 indicates a clear-cut case where the sample features align closely with specific conditions encoded by this rule.

From an interpretability standpoint, these varying activation patterns highlight the transparency of the TSK fuzzy system. Such clarity in rule activation aids practitioners in understanding precisely why certain predictions were made, providing valuable insights for debugging, model improvement, and explaining decisions to stakeholders, establishing trust and confidence in the model. 

If the model uses 3 features, and 3 MFs for each feature, the TSK rules after optimization will be of the form:
\begin{itemize}
    \item (IF $f_1$ is LOW) AND (IF $f_2$ is MEDIUM) AND (IF $f_3$ is HIGH), then $y = b_1 + 0.5f_1 + 0.9f_2 + 0.3f_3 $
    \item (IF $f_1$ is MEDIUM) AND (IF $f_2$ is MEDIUM) AND (IF $f_3$ is HIGH), then $y = b_2 - 0.2f_1 + 0.5f_2 + 0.8f_3 $
    \item (IF $f_1$ is HIGH) AND (IF $f_2$ is LOW) AND (IF $f_3$ is MEDIUM), then $y = b_3 + 0.6f_1 + 0.7f_2 - 0.1f_3 $
\end{itemize}
\dots

\begin{figure}
\centering
\begin{subfigure}{.5\textwidth}
  \centering
  \includegraphics[width=.9\linewidth]{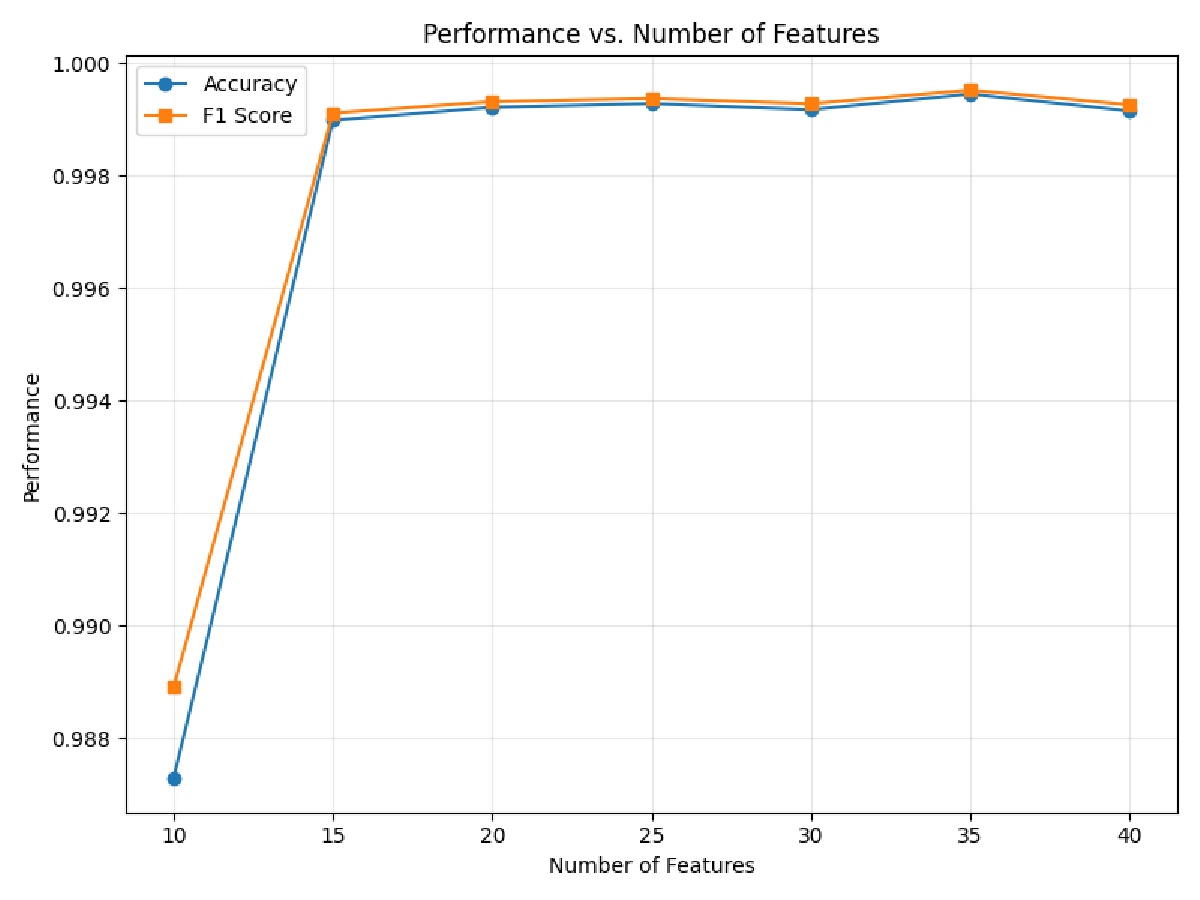}
  \caption{Features vs Performance}
  \label{fig:featuresvperformance}
\end{subfigure}%
\begin{subfigure}{.5\textwidth}
  \centering
  \includegraphics[width=.9\linewidth]{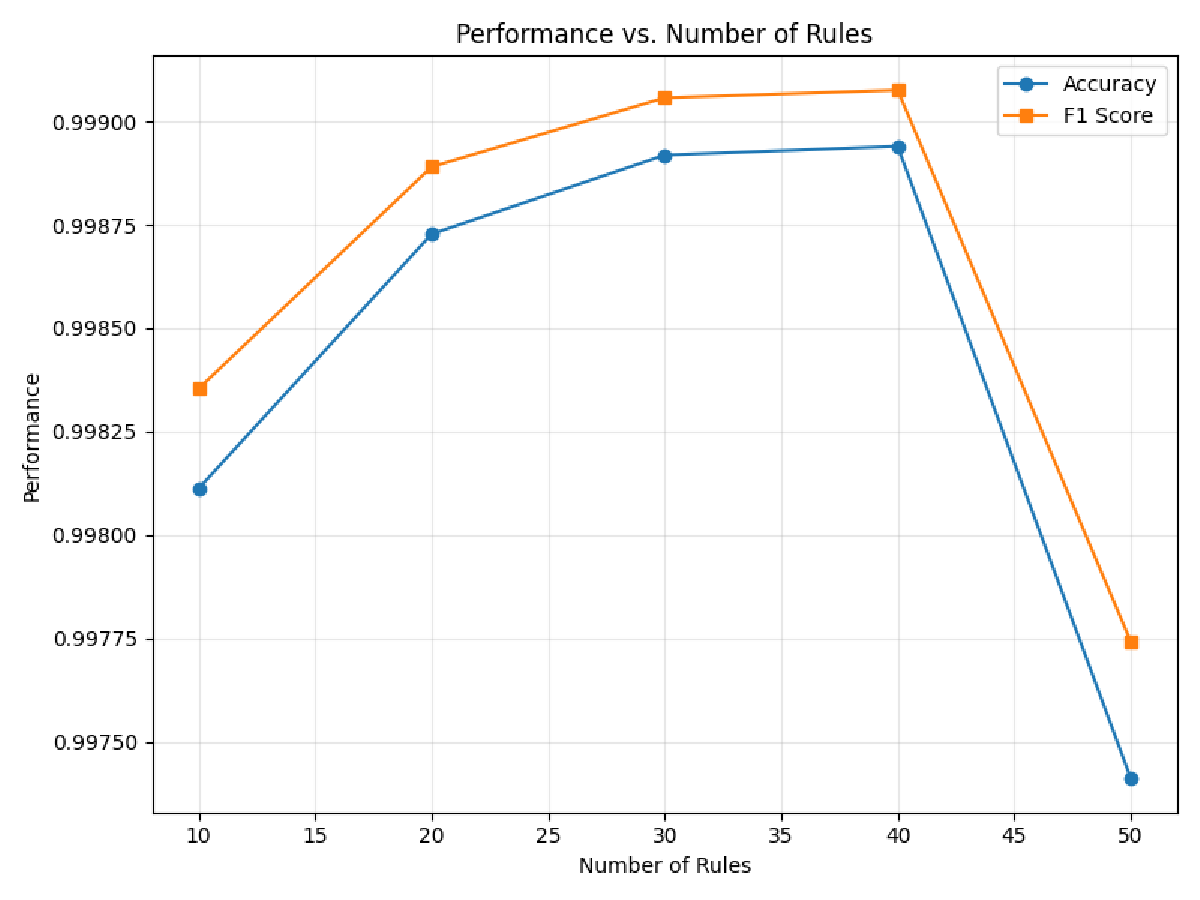}
  \caption{Rules vs Performance}
  \label{fig:rulesvperformance}
\end{subfigure}
\caption{Model Scaling Performance}
\label{fig:scaling_performance}
\end{figure}

Figure \ref{fig:scaling_performance} illustrates that optimal model configurations involve a moderate complexity -- specifically between 15 to 35 features and approximately 20 to 40 rules. Such configurations provide the best balance between model accuracy and interpretability, ensuring high performance without unnecessary complexity or diminishing returns.

\begin{table}[ht]
\centering
\caption{Cross-validation performance comparison of the TSK-Fuzzy model against ensemble classifiers}
\begin{tabular}{lccccc}
\toprule
\textbf{Model} & \textbf{Accuracy} & \textbf{Precision} & \textbf{Recall} & \textbf{F1 Score} & \textbf{AUC} \\
\midrule
\textbf{TSK-Fuzzy (Ours)}      & 0.9995 & 0.9995 & 0.9998 & 0.9996 & 1.0000 \\
Random Forest  & 0.9997 & 0.9996 & 0.9999 & 0.9998 & 1.0000 \\
XGBoost        & 0.9997 & 0.9997 & 0.9998 & 0.9997 & 1.0000 \\
LightGBM     & 0.9998 & 0.9998 & 0.9999 & 0.9998 & 1.0000 \\
\bottomrule
\end{tabular}
\label{tab:benchmark_results}
\end{table}

The proposed gradient-optimized first-order TSK-Fuzzy model achieves strong predictive performance, achieving an average accuracy of 99.95\%, F1 score of 99.96\%, and an AUC of 1.00 across 5-fold cross-validation. When benchmarked against widely-used ensemble classifiers, including Random Forest, XGBoost, and LightGBM, the TSK-Fuzzy system performs competitively, with only marginal differences in predictive performance. While it did not outperform these methods in absolute terms, it offers the unique advantage of interpretability through rule-based fuzzy reasoning. This makes it particularly well-suited for applications that demand transparent and explainable decision-making, such as phishing detection within cybersecurity frameworks. 

\section{Conclusion and Future Work}
In this work, we presented a gradient-optimized first-order Takagi-Sugeno-Kang (TSK) fuzzy inference system for phishing URL detection, integrating interpretable fuzzy logic with the adaptive learning capabilities of gradient descent using the Adam optimizer. Our model demonstrated strong convergence, minimal overfitting, and robust generalization across five cross-validation folds, achieving near-perfect classification performance while maintaining transparency through interpretable rule structures. The ability of the TSK system to visually demonstrate rule activation patterns and evolve membership functions during training underscores its strength in providing intelligible decision boundaries for security-critical applications.

For future work, we aim to incorporate adversarial training strategies to improve the model's resilience against evasion attacks, ensuring robustness in real-world deployment. We also plan to extend this approach to multi-modal phishing detection tasks involving emails, social media messages, and web traffic, broadening the model's utility beyond URL analysis. Finally, investigating real-time deployment feasibility and conducting user studies on the interpretability of the fuzzy rule explanations will help translate the research into actionable, practitioner-focused tools.

\section{Acknowledgements}
The authors extend their sincere gratitude to the members of the AI Bio Lab at the University of Cincinnati for their invaluable discussions and collaborative efforts that facilitated the realization of this work. In particular, the contributions of Magnus Sieverding, Tri Nguyen, and Palak Shah are highly appreciated.

\bibliographystyle{styles/bibtex/spmpsci_unsrt.bst}
\bibliography{bibliography/nafips}

\end{document}